\documentclass{article}

\usepackage{amsmath}
\usepackage{amssymb}
\usepackage{graphicx}

\usepackage[bbgreekl]{mathbbol}

\newtheorem{definition}{Definition}
\newtheorem{proposition}[definition]{Proposition}

\title{Meson mass spectrum using the Cayley-Dickson algebra}
\author{S.~Kuwata, A.~Omoto, S.~Ishihara}
\date{
Faculty of Information Sciences, Hiroshima City University, Asaminami-ku,
Hiroshima 731-3194, Japan}

\begin{document}

\maketitle

\begin{abstract}
From an injective map between the mass of the meson 16-plet and the eigenvalue of the
right multiplication in the Cayley-Dickson algebra,
we obtain the mass formula
as $2 m_{D_s} = m_{\eta_c} + m_{\eta'}$,
which is in excellent agreement with experiment.
\end{abstract}

\section{Introduction}

In the standard quark model, mesons are bound states of a quark and anti-quark.
Due to the difference between quark masses, the SU($N$) flavor symmetry is broken.
Concerning the meson mass formula,
the Gell-Mann--Okubo relation~\cite{gellmann,okubo} is well known.
For $N=3$, the Gell-Mann--Okubo formula for the pseudoscalar mesons is given by
$3 m_\eta \simeq 4 m_K - m_\pi$, assuming no singlet-octet mixing, which, however, cannot be neglected due to the SU(3) symmetry breaking.
It is difficult to estimate the singlet-octet mixing from the improvement of this formula (or other meson mass formula based on the dual resonance model~\cite{frampton} in connection with string theory) by taking account of the quark-quark interaction mediated by the gluon exchange~\cite{donoghue}.
It may remain difficult to obtain the singlet-octet mixing, even if the SU(4) meson 16-plet and SU(5) 25-plet are taken into account~\cite{burakovsky,burakovsky2}. For the meson 16-plet, for example, the model such that the SU(4) symmetry is broken but the SU(3) symmetry is exact, leads to a simple mass formula of the form $12 m_{\bar{D}}^{2} = 5 m_{c\bar{c}}^2 + 7 m_0^2$~\cite{burakovsky}, where $m_{\bar{D}}$ represents the average mass of $c \bar{u}, c\bar{d}$, and $c\bar{s}$; and $m_0$ stands for the average of the meson octet masses; due to the assumption of the exact SU(3) symmetry, the mass relation between the physical states $\eta$ and $\eta'$ cannot be obtained.

The aim of this paper is obtain a simple but exact mass formula
by relating the meson $n$-plet  (denoted by $\mathbb n$ for brevity) to a vector space where some appropriate algebra is given.
The basic technique was developed in Ref.~\cite{alternative}, where the meson octet can be identified with a certain algebra.
Here by algebra $\mathcal A$, we mean a hypercomplex system, that is, a vector space $V$ with a given multiplication $L_x : V \rightarrow V$ (with $x \in V$).
Denote by $m_i$ (for $i=1, \ldots, n$) the mass of the meson $n$-plet, and
$\lambda_i$ (for $i=1, \ldots, n$) the eigenvalue of $L_x$ with $x \in V / V_0$, where $V_0 \subset V$ represents the subspace of $V$ such that $\dim (V / V_0) = n$ (so that $\dim V$ should be a multiple of $n$).
The reason of dividing $V$ by $V_0$ is that we want to obtain a bijective map between the meson mass and the eigenvalues of $L_x$ (in an actual case, $L_x$ is replaced by its relative).
Suppose that there is a bijective map $\phi: M \rightarrow \Lambda$,
where $M = \cup_i \{ m_i \}$ and $\Lambda = \cup_i \{ \lambda_i \}$.
Then it is found that there is a bijective map $\tilde{\phi}: \mathbb n  \rightarrow V / V_0$, due to the existence of a bijective maps $f: \mathbb n \rightarrow M$ and $g: \Lambda \rightarrow V / V_0$, that is,
$\tilde{\phi} = g \circ \phi \circ f$ [see Eq.~(\ref{eq:diagram})].
Conversely, if the algebra $\mathcal A$ is given (this means that the maps $\tilde{\phi}$ and $g$ are given), then the eigenvalue of (the relative of) $L_x$ is related to the meson mass through the relation of $\phi \circ f = g^{-1} \circ \tilde{\phi}  \; (= {\rm given})$.

\begin{align}
\begin{array}{r@{}ccc@{}l}
& {\mathbb n} &  \overset{\tilde{\phi}}{\longrightarrow}  & V / V_0 &  \\
{\scriptstyle f}  & \downarrow & & \uparrow  &  {\scriptstyle g} \\
& M &  \overset{\phi}{\longrightarrow} & \Lambda
\end{array}
\label{eq:diagram}
\end{align}
As an example, 
consider the meson octet $\mathbb 8$, which may be composed as
\begin{align}
\mathbb 8 &= \mathbb 1 \oplus \mathbb  1 \oplus \mathbb 2 \oplus \mathbb 4 \notag \\
&= \eta \oplus \pi^0 \oplus ( \pi^+, \pi^- ) \oplus (K^+, K^-; K^0, \bar{K}^0 ).
\label{eq:octet}
\end{align}
Although there may be many algebras such that the map
$\tilde{\phi}: \mathbb 8 \rightarrow V / V_0$
is injective, we have already chosen in Ref.~\cite{alternative} the Cayley-Dickson algebra with
$V = \mathbb R^{64}$ and $V_0 = \mathbb R^8 $.

The reason of adopting the Cayley-Dickson algebra is as follows.
Originally, the flavor SU$(N)$ symmetry breaking is responsible for the meson mass difference.
Recall that SU$(N)$ is one of the compact simple Lie groups, which are categorized
into two types: classical and exceptional ones. The classical type represents the
isometry transformation in the vector space over the real number $\mathbb R \; (=\mathbb A_0)$, the
complex number $\mathbb C \; (= \mathbb A_1)$, and the Hamilton number (quaternion) $\mathbb H \; (= \mathbb A_2)$. 
The isometry over the Cayley number (octonion) $\mathbb O \; (= \mathbb A_3)$ is not a
simple Lie group of a classical type, due to the lack of the associativity.
However, due to the alternativity of $\mathbb O$,
$\mathbb O$ is still related to the simple Lie group of an exceptional type
as~\cite{baez}
G$_2 \cong {\rm Aut} (\mathbb O)$,
F$_4 \cong {\rm Isom} ( \mathbb O \mathbb P^2)$,
E$_6 \cong {\rm Isom}((\mathbb C \otimes \mathbb O) \mathbb P^2)$,
E$_7 \cong {\rm Isom}((\mathbb H \otimes \mathbb O) \mathbb P^2)$,
E$_8 \cong {\rm Isom}((\mathbb O \otimes \mathbb O) \mathbb P^2)$,
where ${\rm Aut}$, ${\rm Isom}$, and $\mathbb K \mathbb P^2$ represent the
automorphism, isometry, and the projective plane over a (skew) field $\mathbb K$,
respectively.
Recall also that $\mathbb A_n$ is the $2^n$-dimensional Cayley-Dickson algebra over $\mathbb R$.
The above mathematical facts implies that if part of the symmetry under simple Lie group is not broken,
a subgroup of the isometry group over $\mathbb A_n$ (with $n \leq 3$) may still survive.
Noticing that $\mathbb A_n$ is a subalgebra of $\mathbb A_{n'}$ (for $n < n'$),
we may naturally choose $\mathbb A_n$ (with $n \geq 4$) as an algebra by which the Lie group symmetry breaking in the sense of Eq.~(\ref{eq:diagram})
(actually, $\mathbb A_6 = \mathbb R^{64}$ is chosen for the meson octet $\mathbb 8$).

Furthermore, it should be mentioned why we deal with the meson $n$-plet, rather than the baryon $n$-plet.
The reason is simple.
Consider, for example, the baryon octet, which is composed of $\Lambda, \Sigma^0, \Sigma^+, \Sigma^-, p, n, \Xi^0, \Xi^-$.
Difference from the meson octet, the baryon octet cannot be identified with $\mathbb A_6 / \mathbb R^8$ because the baryon octet cannot be decomposed into $\mathbb 1 \oplus \mathbb 1 \oplus \mathbb 2 \oplus \mathbb 4$, where the masses in $\mathbb 4$ are doubly degenerate.

The outline of this paper is as follows.
In Sec.~2, we briefly review the basic property of the Cayley-Dickson algebra, where the eigenvalues of (the relative of) $L_x$ are given.
In Sec.~3, 
we apply to the pseudoscalar meson 16-plet the analogous map $\tilde{\phi}$ that can be applied to the meson octet, to finally obtain the mass formula $2 m_{D_s} = m_{\eta_c} + m_{\eta'}$, which is well verified by experiment.
In Sec.~4, we make further application to the vector meson 16-plet and to the meson 25-plet.
In Sec.~5, we give summary.

\section{Cayley-Dickson algebra}

\label{sec:cayley}

In this section, we briefly review the basic property of the Cayley-Dickson
algebra. The Cayley-Dickson algebra $\mathbb A_n$ over the real number $\mathbb R$
represents the algebra structure on $\mathbb R^{2^n}$, which is given inductively.
Let $x=(x_1, x_2), y = (y_1,y_2)$ be in
$\mathbb R^{2^n} = \mathbb R^{2^{n-1}} \times \mathbb R^{2^{n-1}}$.
Then the multiplication $xy$ is given by
\begin{align*}
x y = (x_1 y_1 - \bar{y}_2 x_2, y_2 x_1 + x_2 \bar{y}_1),  \quad \mbox{with }
\bar{x} = ( \bar{x}_1, - x_2).
\end{align*}
For $n \leq 3$, $\mathbb A_n$ corresponds to
\begin{align*}
\mathbb A_0 = \mathbb R, \;\mathbb A_1 = \mathbb C, \; \mathbb A_2 = \mathbb H, \;
\mathbb A_3 = \mathbb O,
\end{align*}
which and only which are normed division algebras. The basic property of $\mathbb A_n$ is summarized in Table~\ref{t:basic}.
\begin{table}[b]
\caption{Basic property of $\mathbb A_n$~\cite{moreno}, where the commutator and associator are given by $[x, \, y] = xy - yx$, $[x, \, y, \, z] = (xy)z - x (yz)$, respectively.}
\begin{center}
\begin{tabular}{lll}
\hline
$n$ & Property & Identity \\
\hline
$0$ & Self conjugate &$x = \bar{x}$ \\
$0,1$ & Commutative & $[x, \, y ] = 0$ \\
$0,1,2$ & Associative & $[x, \, y, \, z] = 0$ \\
$0,1,2,3$ & Alternative & $[x, \, x, \, y ] = 0$ \\
All & Flexible & $[x, \, y, \, x] = 0$ \\
\hline
\end{tabular}
\end{center}
\label{t:basic}
\end{table}
The Euclidean norm and inner product are given by $\| x \|^2 = x \bar{x} = \bar{x} x$ and
$\langle x, y \rangle = \frac{1}{2} ( x \bar{y} + y \bar{x} )$, respectively.
Due to the flexibility of $\mathbb A_n$, one obtains for all $x, y, z \in \mathbb A_n$ the idenities~\cite{moreno}
\begin{align}
\langle x, y z \rangle = \langle x \bar{z}, y \rangle = \langle \bar{y} x , z \rangle.
\label{eq:inner}
\end{align}

For further analysis of  the algebra structure of $\mathbb A_n$, it is convenient to define
the left and right multiplications
$L_x, R_x: \mathbb A_n \rightarrow \mathbb A_n$ by
\begin{align*}
L_x (y) = xy, \quad R_x (y) = yx
\end{align*}
for $x \in \mathbb A_n$ fixed.
If one tries to decompose the vector space $\mathbb R^{2^n}$ into the eigenspaces of $L_x$, one encounters an obstacle; the eigenvalue of $L_x$ is not necessarily given by a real number so that the eigenspace cannot be given by a real number.
To remove the obstacle, we deal with the eigenvalue of $N_x : = L_{\bar{x}} L_x$,
instead of the eigenvalue of $L_x$ itself.
Since $N_x$ is (real) symmetric, that is, $\langle y, N_x (z) \rangle = \langle y, \bar{x} (xz) \rangle = \langle xy, xz \rangle = \langle xz, xy \rangle = \langle z, N_x (y) \rangle$ by Eq.~(\ref{eq:inner}),
all the eigenvalues of $N_x$ turn out  to be real numbers, so that
the vector space $\mathbb R^{2^n}$ can be decomposed into the eigenspaces of $N_x$.

Denote by $S_n$ the set of the eigenvalues of $N_x - \| x \|^2$ for
$x=(x_1,x_2) \in \mathbb A_n = \mathbb A_{n-1} \times \mathbb A_{n-1}$.
For $n \leq 4$, it is relatively easy to calculate $S_n$ as
\begin{align*}
S_i &= \{ \underbrace{0, \ldots, 0}_{2^i \mbox{ \scriptsize times}} \} \quad
(\mbox{for } i=0,1,2,3), \\ 
S_4 &= \{ \underbrace{0, \ldots, 0}_{8 \mbox{ \scriptsize times}} ; \,
\underbrace{\pm \Delta^2, \ldots, \pm \Delta^2}_{4 \mbox{ \scriptsize times}} \},
\end{align*}
where $\Delta  = 2 | {\bf x}_1 \times {\bf x}_2 | := 2 \sqrt{ \| {\bf x}_1 \|^2 \| {\bf
x}_2 \|^2 - \langle {\bf x}_1, {\bf x}_2 \rangle^2 }$, with the bold face letter ${\bf
x}$ representing the imaginary part of $x$, that is, ${\bf x} = x - {\rm Re} (x) =
\frac{1}{2} ( x - \bar{x} )$.
For $n \geq 5$, the calculation of $S_n$ turns out to be so difficult that it may
not be useful for a physical application.
Before proceeding further, it should be noticed that $S_n$ satisfies the following inclusion relation:
\begin{align*}
\begin{cases}
S_{n-1} \subset S_n &  (\mbox{for } n \leq 4), \\
S_{n-1} \not{\! \! \subset} \; S_n &  (\mbox{for } n \geq 5).
\end{cases}
\end{align*}
The inclusion relation of $S_{n-1} \subset S_n$ implies that 
half of the eigenspaces of $N_x$ for $x \in \mathbb A_n$ are given by the eigenspaces of $N_x$ with $x$ restricted to $\mathbb A_{n-1}$, which is the subspace of $\mathbb A_n$.
In the present study, we extend to the map applied to the meson octet to the meson 16-plet, so that the inclusion relation of $S_{n-1} \subset S_n$ should be satisfied. Otherwise, the original map $\tilde{\phi}:\mathbb 8 \rightarrow \mathbb A_6 / \mathbb R^8$, in itself, would be violated when applied to the meson 16-plet.

Now we obtain the necessary and sufficient condition for $S_{n-1}
\subset S_n$.
Recall that for $n \geq 5$, $\mathbb A_n$ is not given by a pair of alternative elements in $\mathbb
A_{n-1}$ (which is refered to as alternative entries, for short), where $S_{n-1} \subset S_n$ does not hold in general.
Thus it is necessary for $S_{n-1} \subset S_n$ that an element in $\mathbb A_n$ is given by alternative entries.
Here, the alternative element is given by the following definition~\cite{moreno}:
\begin{definition}
$a$ in $\mathbb A_n$ is an alternative element, if $[a, \, a, \, x] = 0$ holds for
all $x$ in $\mathbb A_n$.
\end{definition}
Recall that if all the elements in $\mathbb A_n$ are alternative, then we simply call $\mathbb A_n$ alternative.
Fortunately, it is sufficient for $S_{n-1} \subset S_n$ that an element in $\mathbb A_n$ is given by alternative entries:
\begin{proposition}
If an element in $\mathbb A_n$ is given by a pair of alternative elements in $\mathbb A_{n-1}$,
then it follows that $S_{n-1} \subset S_n$.
\end{proposition}
The proof, however, is somewhat complicated, and is referred to our previous work~\cite{alternative}.

Once an element $x \in \mathbb A_n$ (for $n \geq 3$) is given by alternative entries,
the eigenpolynomial for $N_x$ turns out to be an even function with quadruple degeneracy. In this case, $S_n$ can be written as
\begin{align*}
S_n = \bigcup_{i=1}^4 \left( \tilde{S}_n  \cup (-\tilde{S}_n) \right) \quad (\mbox{for } n \geq 3)
\end{align*}
where the element in the set $\tilde{S}_n$ represents the quadruply degenerated non-negative eigenvalues of $S_n$.
Under an appropriate parameterization, 
$\tilde{S}_n$ is given by~\cite{alternative}
\begin{align}
\begin{split}
\tilde{S}_3 &= \{ 0 \} , \\
\tilde{S}_4 \setminus \tilde{S}_3 &= \{ \Delta^2 \}, \\
\tilde{S}_5 \setminus \tilde{S}_4 &= \{ \Delta_+^2, \Delta_-^2  \}, \\
\tilde{S}_6 \setminus \tilde{S}_5 &= \{ \Delta_{+-}^2, \Delta_{-+}^2; \Delta_{++}^2, \Delta_{--}^2 \}, \\
\tilde{S}_7 \setminus \tilde{S}_6 &= \{ \Delta_{+--}^2, \Delta_{-++}^2; \Delta_{+-+}^2, \Delta_{-+-}^2; 
\Delta_{++-}^2, \Delta_{--+}^2; \Delta_{+++}^2, \Delta_{---}^2 \},
\end{split}
\label{eq:Sn}
\end{align}
and so on, where
$\Delta_{\underbrace{\scriptstyle \pm \cdots \pm}_{k \; \rm times}} = \Delta \cdot
\cos (\pm \theta_{k} \pm\theta_{k-1} \cdots \pm \theta_1)$.
The parameters $\theta_i$ ($i=1,2,\ldots, k$) are introduced by the requirement from the alternative entries.
Thus for $n \leq 4$, there is no such parameter in ${S}_n$, because all the elements in $\mathbb A_n$ are given by alternative entries.

\section{Mass formula}

In this section, we obtain a simple mass formula as
$2  m_{D_s} = m_{\eta_c} + m _{\eta'}$
by extending the injective map $\tilde{\phi}$ to the meson 16-plet.
Comparing Eq.~(\ref{eq:octet}) and $\tilde{S}_6$ in Eq.~(\ref{eq:Sn}), we readily find that there is an injective map $\phi \circ f: \mathbb 8 \rightarrow \tilde{S}_6$, so that there is a one-to-one correspondence between the meson octet $\mathbb 8$ and $\mathbb A_6$ as
\begin{align}
\tilde{\phi} : \mathbb 8 \rightarrow \mathbb A_6 / \mathbb R^8.
\label{eq:8}
\end{align}
The reason of dividing $\mathbb A_6$ by $\mathbb R^8$ is due to the inclusion map $i$
\begin{align*}
i: \tilde{S}_n \hookrightarrow S_n.
\end{align*}
To make the correspondence concrete, let  $\theta: \mathbb n \rightarrow \mathbb R$ be defined as $\theta = j \circ \phi \circ f$ with $j: \mathbb R \rightarrow \mathbb R$ by $j (x) = \arccos \sqrt{x/\Delta^2}$. Then the correspondence between the meson octet $\mathbb 8$ and the element in $\tilde{S}_6$ is summarized as in the left-hand column of Table~\ref{t:corresp}, from which it is found that the parameter $\theta_1$ represents the difference between $u$ and $d$ quarks, and the parameter $\theta_2$ the difference between $s$ and $u$ (or $d$) quarks. 
\begin{table}[b]
\caption{Correspondence between the meson 16-plet and the elements in $\tilde{S}_7$, where $\theta$ is a parameter representing an element in $\tilde{S}_n$ such that $x = \Delta^2 \cos^2 \theta$ for $x \in \tilde{S}_n$.}

\begin{tabular}{ccp{1cm}cc}
\hline
$\mathbb 8$ & $\theta$ & & $\mathbb{16} \setminus \mathbb 8$ & $\theta$ \\
\hline
$\pi^0$ & $0$ & & $D^0, \bar{D}^0$ & $\pm (\theta_3 - \theta_2 - \theta_1)$ \\
$\pi^\pm$ & $\pm \theta_1 $ & & $D^\pm$ & $\pm (\theta_3 - \theta_2 + \theta_1)$ \\
$K^\pm$ & $\pm (\theta_2 - \theta_1) $ & & $D_s^\pm$ & $\pm (\theta_3 + \theta_2 - \theta_1)$ \\
$K^0, \bar{K}^0$ & $\pm (\theta_2 + \theta_1) $ & & $\frac{1}{\sqrt{2}} (\eta_c \pm \eta')$ & $\pm (\theta_3 + \theta_2 + \theta_1)$ \\
\hline
\end{tabular}
\label{t:corresp}
\end{table}
At the present stage, we have no mass relation unless the map $\phi: M \rightarrow \tilde{S}_6$ is specified.

\begin{figure}[bth]
%
\begin{center}
\unitlength 0.71mm 
\begin{picture}(150,65)(0,15)
%
\put(17,47){\circle*{3}}
\put(32,37){\circle*{3}}
\put(37,57){\circle*{3}}
\put(77,57){\circle*{3}}
\put(72,37){\circle*{3}}
\put(92,47){\circle*{3}}
{\thicklines
\put(17,47){\line(3,-2){15}}
\put(17,47){\line(2,1){20}}
\put(32,37){\line(1,0){40}}
\put(37,57){\line(1,0){40}}
\put(72,37){\line(2,1){20}}
\put(92,47){\line(-3,2){15}}
}
\put(37,72){\circle*{3}}
\put(52,62){\circle*{3}}
\put(77,72){\circle*{3}}
{\thicklines
\put(37,72){\line(1,0){40}}
\put(37,72){\line(3,-2){15}}
\put(52,62){\line(5,2){25}}
}
\put(32,22){\circle*{3}}
\put(57,32){\circle*{3}}
\put(72,22){\circle*{3}}
{\thicklines
\put(32,22){\line(1,0){40}}
\put(32,22){\line(5,2){25}}
\put(57,32){\line(3,-2){15}}
}
%
\put(17,47){\line(4,5){20}}
\put(32,37){\line(4,5){20}}
\put(72,37){\line(-4,5){20}}
\put(92,47){\line(-3,5){15}}
\put(37,57){\line(0,1){15}}
\put(77,57){\line(0,1){15}}
\put(17,47){\line(3,-5){15}}
\put(32,37){\line(0,-1){15}}
\put(72,37){\line(0,-1){15}}
\put(92,47){\line(-4,-5){20}}
\put(57,32){\line(4,5){4.0}} 
\put(57,32){\line(-4,5){4.0}} 

\put(52,45.5){\circle*{2}}
\put(53,48.5){\circle*{2}}
\put(55,45.5){\circle*{2}}
\put(56,48.5){\circle*{2}}

\put(120,47){\vector(-1,-2){10}}
\put(120,47){\vector(1,0){20}}
\put(120,47){\vector(0,1){25}}
\put(135,40){$\sf I_3$}
\put(115,28){$\sf S$}
\put(122.5,65){$\sf C$}

\put(49,49){$\pi^0$}
\put(56,50){$\eta$}
\put(46.5,41.5){$\eta'$}
\put(53.5,41.5){$\eta_c (c \bar{c})$}
\put(10,47){$\pi^-$}  \put(21,45){$d \bar{u}$}
\put(94,47){$\pi^+$}  \put(80,44.6){$u \bar{d}$}

\put(33,31){$K^+$}  \put(36,38.5){$u \bar{s}$}
\put(69,59){$K^-$}   \put(73,51){$ s \bar{u}$}
\put(72.5,31.5){${K}^0$}  \put(70,40){$d \bar{s}$}
\put(38,59){$\bar{K}^0$}   \put(35,51){$ s \bar{d}$}

\put(27.5,71){$D^0$}  \put(34.5,75){$c \bar{u}$}
\put(79,71){$D^+$}  \put(73,75){$c \bar{d}$}
\put(55,59){$D_s^+$}  \put(51,65){$c \bar{s}$}
\put(22,19){$D^-$}  \put(29,15){$d \bar{c}$}
\put(46,32.2){$D_s^-$}  \put(52.5,26){$s \bar{c}$}
\put(74,19){$\bar{D}^0$}  \put(68,16){$u \bar{c}$}

\end{picture}

\end{center}
\caption{Meson 16-plet,
where {\sf S} represents the strangeness and
{\sf I}$_3$ is the $z$-component isospin {\sf I}.}
\label{f:16plet}
\end{figure}
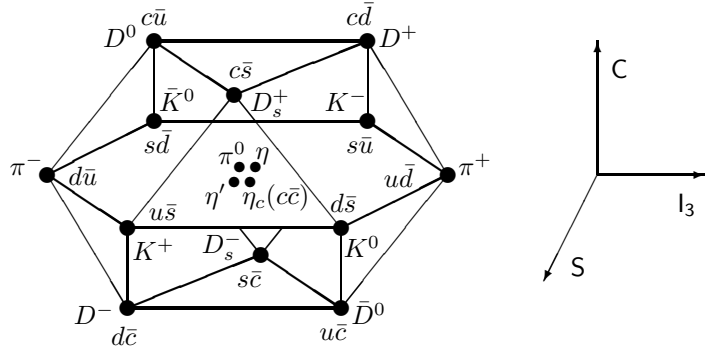

If we take account of the meson 16-plet, which is constructed from the SU(4) flavor system, it is expected that an analogous map to Eq.~(\ref{eq:8}) should hold: $\tilde{\phi} : \mathbb{16} \rightarrow  \mathbb A_7 / \mathbb R^8$.
If so, we have an injective map $\phi \circ f : \mathbb{16} \setminus \mathbb 8  \rightarrow \tilde{S}_7 \setminus \tilde{S}_6$, where $\mathbb{16} \setminus \mathbb 8$ is given by (see Fig.~\ref{f:16plet})
\begin{align}
\mathbb{16} \setminus \mathbb 8 = (D^0, \bar{D}^0) \oplus  (D^+, D^-) \oplus (D_s^+, D_s^-) \oplus \eta_c \oplus \eta'.
\label{eq:16/8}
\end{align}

At first glance, there seems to be no such injective map $\phi \circ f$.
This is because while the elements in $\tilde{S}_7 \setminus \tilde{S}_6$ are doubly degnerate, the masses in $\mathbb{16} \setminus \mathbb 8$ are not necessarily so; the mass of $\eta_c$ is not equal to that of $\eta'$.
However, a slight modification of the basis of the meson field of $\eta_c$ and $\eta'$ will lead to a desirable result.
Notice that among the mesons in $\mathbb{16} \setminus \mathbb 8$,
$\eta_c$ and $\eta'$ are the (only) mesons that have the same
quantum numbers: zero electric charge, zero isospin, zero strangeness, and zero charm.
Thus, it is physically possible to consider a superposed state of $\eta_c$ and $\eta'$.
To make the two (orthogonal) superposed mesons have the same mass, $\eta_c \oplus \eta'$ in Eq.~(\ref{eq:16/8}) should be transformed to a ``doublet'' as
\begin{align*}
\eta_c \oplus \eta' \longrightarrow (\eta_+, \eta_-),
\end{align*}
where $\eta_\pm = \frac{1}{\sqrt{2}} (\eta_c \pm \eta')$.
In this case, the $\theta$-assignment for $\mathbb{16} \setminus \mathbb 8$ is summarized in the right-hand column of Table~\ref{t:corresp}.

To obtain a mass relation, it should be recalled that the parameter $\theta_1$ represents the mass difference between the $u$ and $d$ quarks. Considering the empirical relation of $m_u \approx m_d$, we find that $\theta_1 \approx 0$ as long as the map $\theta: \mathbb n \rightarrow \mathbb R$ is continuous.
In this case, we obtain
$m_{D_s} \approx m_{\eta_\pm}$ from $\theta ({D_s}) \approx \theta ({\eta_\pm})$, that is
\begin{align}
2 m_{D_s} \approx m_{\eta_c} + m_{\eta'}.
\label{formula}
\end{align}
The recent experimental value of the meson mass (see Table~\ref{t:data}) leads to
\begin{align*}
\frac{ m_{\eta_c} + m_{\eta'} }{ 2 m_{D_s} } = 1.00033 \pm 0.00035,
\end{align*}
which indicates the validity of the relation Eq.~(\ref{formula}). In evaluating the standard deviation, we have assumed that $m_{\eta_c}, m_{\eta'}$, and $m_{D_s}$ are independent variables.
\begin{table}[bth]
\caption{Recent experimental value of the meson mass~\cite{meson} .}
\begin{center}
\begin{tabular}{cc}
\hline
Meson & Mass (MeV) \\
\hline
$\eta_c$ & $2980.5 \pm 1.2$ \\
$\eta'$ & \, $957.78 \pm 0.06$ \\
$D_s$ & $1968.49 \pm 0.34$ \\
\hline
\end{tabular}
\end{center}
\label{t:data}
\end{table}

At the end of this section, we discuss the usefulness of Eq.~(\ref{formula}) as an $\eta'$-including mass formula.
In the standard quark model, however, it is not so easy a task to relate $m_{\eta'}$ to other
meson masses, due to several reasons.
One is that $\eta'$ is not a pure SU(3) singlet $\eta_1$, but a mixture with one of the SU(3) octet, $\eta_8$, through the relation
\begin{align}
\begin{pmatrix}
\eta' \\ \eta
\end{pmatrix} =
\begin{pmatrix}
\cos \vartheta & \sin \vartheta \\
- \sin \vartheta & \cos \vartheta
\end{pmatrix}
\begin{pmatrix}
\eta_1 \\ \eta_8
\end{pmatrix},
\label{eq:mix}
\end{align}
where $\eta_1 = \frac{1}{\sqrt{3}} ( u \bar{u} + d \bar{d} + s \bar{s} )$ and
$\eta_8 = \frac{1}{\sqrt{6}} ( u \bar{u} + d \bar{d} - 2 s \bar{s} )$.
To eliminate the mixing angle $\vartheta$ from the theory,
one of the orthodox methods is to introduce the pion and kaon masses, with the result known as the Schwinger relation~\cite{schwinger}.
However, the Schwinger relation,
which holds under the assumption of the ``ideal mixing'' $\eta' = s \bar{s}$ and $\eta = \frac{1}{\sqrt{2}} (u \bar{u} + d \bar{d})$, is not satisfactory for the pseudoscalar mesons, due to the large deviation from the ideal mixing.
For the pseudoscalar mesons,
the interaction between the different flavor (such as $u \bar{u}
\leftrightarrow s \bar{s}$) cannot be neglected,
so that the mass caused by this interaction is comparable to the constituent
quark mass~\cite{feldmann}.
Taking these things into account, we find it somewhat marvelous that $m_{\eta'}$
satisfies so simple a relation like Eq.~(\ref{formula}).

\section{Further application}

In this section, we apply an analogous map $\tilde{\phi}$ to the vector meson 16-plet and to the meson 25- plet.
However, it is found that further application to the vector 16-plet causes a delicate problem and that the application to the 25-plet is not viable as follows.

First, we deal with the case of  vector meson 16-plet, where
$\rho, K^\ast, \phi, \omega, J/\psi, D^\ast, D_s^\ast$ take the place of $\pi, K, \eta, \eta', \eta_c, D, D_s$, respectively. Suppose that the map $\tilde{\phi}$ can be applied to the vector meson 16-plet,
the relation of Eq.~(\ref{formula}) might be replaced by $2 m_{D_s^\ast} \approx m_{J/\psi} + m_{\omega}$, which, however, is not satisfactory compared to Eq.~(\ref{formula}).
The failure of the relation of $2 m_{D_s^\ast} \approx m_{J/\psi} + m_{\omega}$ can be interpreted as follows.
In deriving Eq.~(\ref{formula}), it should be recalled that we assume that the injective map $\theta: \mathbb{n} \rightarrow \mathbb R$ is continuous. This implies that $\cos \theta$ should decrease monotonically with respect to the meson mass, that is, $\cos \theta (\varphi) \gtrless \cos \theta (\varphi') \Longleftrightarrow m_\varphi\lessgtr m_{\varphi'}$ for $\varphi, \varphi' \in \mathbb{n}$ (see Table~\ref{t:corresp}, where $m_{\pi^0}$ is the smallest).
Thus the condition of $m_{\rho^0} < m_{\rho^\pm}$ is necessary for the relation of
$2 m_{D_s^\ast} \approx m_{J/\psi} + m_{\omega}$.
Different from the pseudoscalar meson, it is quite a delicate problem to determine the sign of $\Delta m_{\rho} \; (\equiv m_{\rho^0} - m_{\rho^\pm})$.  
The Particle Data Group gives the values of $\Delta m_\rho = -0.7 \pm 0.8 \, {\rm MeV}$~\cite{meson}, while some theoretical considerations indicate that $- 0.4 \, {\rm MeV} < \Delta m_{\rho} < 0.7 \, {\rm MeV}$~\cite{bijnens}, $\Delta m_{\rho} = -0.02 \pm 0.02 \, {\rm MeV}$~\cite{feuillat},
$\Delta m_{\rho} = 0.62 \, {\rm MeV}$~\cite{gao}, and $\Delta m_{\rho} \sim 1 \, {\rm MeV}$~\cite{djukanovic}.
As long as $\Delta m_{\rho}$ is positive, it is not necessary to hold a relation of $2 m_{D_s^\ast} \approx m_{J/\psi} + m_{\omega}$ in the sense above. Conversely, the failure of the relation of $2 m_{D_s^\ast} \approx m_{J/\psi} + m_{\omega}$ suggests the relation of $m_{\rho^0} > m_{\rho^\pm}$.

Finally, we deal with the meson 25-plet, where $b$ quark is taken into account. In this case, we cannot single out an enlarged algebra $\mathbb A_8 / \mathbb R^8$, due to $\dim (\mathbb A_8 / \mathbb R^8) = 32 \; ( > 25)$. Thus, we can only make a decomposition as
$\mathbb{25} = \mathbb{16} \oplus \mathbb 8_b \oplus \mathbb 1_b$, where $\mathbb 8_b$ and $\mathbb 1_b \; (= b \bar{b})$ represent the $b$-quark related octet and singlet, respectively.
Although the $\mathbb 8_b$ can be identified with $(\mathbb A_7 \setminus \mathbb A_6 ) / \mathbb R^8$ due to the injective map $\phi \circ f : \mathbb 8_b \rightarrow \tilde{S}_7 \setminus \tilde{S}_6$ [this should be contrasted with the injective map $\phi \circ f: \mathbb 8 \rightarrow \tilde{S}_6$ for the original meson octet $\mathbb 8$ in Eq.~(\ref{eq:octet})], no new mass relation is obtainable as in the case of the original $\mathbb 8$.

\section{Summary}

So far, we have obtained the mass formula Eq.~(\ref{formula}) by identifying the pseudoscalar meson 16-plet with $\mathbb A_7 / \mathbb R^8$ through the injective map $\tilde{\phi}: \mathbb{16} \rightarrow \mathbb A_7 / \mathbb R^8$. The point is that $\eta_c$ and $\eta'$ can be mixed to form a ``doublet''  $(\eta_+, \eta_-)$ so as to have the same mass. This mixture is possible because $\eta_c$ and $\eta'$ have the same quantum numbers as charge, isospin, strangeness, and charm. The resultant mass formula is well verified by experiment.
The application of an analogous map $\tilde{\phi}$ to the vector meson 16-plet brings about quite a delicate problem in connection with the sign of $m_{\rho^0} - m_{\rho^\pm}$.
Further application to the meson 25-plet is not viable due to the lack of an enlarged algebra.

\section*{Acknowledgments}
The authors are indebted to K.~Terada and M.~Terauchi for their stimulating
discussion.

\end{document}